\documentclass[conference]{IEEEtran}

\IEEEoverridecommandlockouts
\usepackage{cite}
\usepackage{amsmath,amssymb,amsfonts}

\usepackage{graphicx}
\usepackage{textcomp}
\usepackage{xcolor}

\usepackage{bm}
\usepackage{algorithm}
\usepackage{float} 
\usepackage{caption}
\usepackage{subfigure}
\usepackage{algorithm}
\usepackage[noend]{algpseudocode}
\usepackage[normalem]{ulem}

\def\BibTeX{{\rm B\kern-.05em{\sc i\kern-.025em b}\kern-.08em
    T\kern-.1667em\lower.7ex\hbox{E}\kern-.125emX}}

\begin{document}

\title{Targeted False Data Injection Attack against DC State Estimation without Line Parameters

\thanks{The authors are with the Department of Electrical and Computer Engineering, McGill University, Montreal, QC H3A 0G4, Canada. (e-mail: mingqiu.du@mail.mcgill.ca, georgia.pierrou@mail.mcgill.ca, xiaozhe.wang2@mcgill.ca).

This work is supported by Fonds de recherche du Qu\'{e}bec--Nature et technologies (FRQNT) under Grant FRQ-NT PR-253686 and Natural Sciences and Engineering Research Council (NSERC) Discovery Grant RGPIN-2016-04570.}}

\author{Mingqiu Du,~\IEEEmembership{Student Member,~IEEE}, \textcolor{black}{Georgia Pierrou,~\IEEEmembership{Student Member,~IEEE} }, Xiaozhe Wang,~\IEEEmembership{Senior Member,~IEEE}}
\maketitle
\vspace{-30pt}
\begin{abstract}
A novel false data injection attack (FDIA) model against DC state estimation is proposed, which requires no network parameters and exploits only limited phasor measurement unit (PMU) data. The proposed FDIA model can target specific states and launch large deviation attacks \textcolor{black}{using estimated line parameters}. Sufficient conditions for the proposed method are also presented. Different attack vectors are studied in the IEEE 39-bus system, showing that the proposed FDIA method can successfully bypass the bad data detection (BDD) with high success rates \textcolor{black}{of up to 95.3\%}.

\end{abstract}

\begin{IEEEkeywords}
DC state estimation, false data injection attacks, stochastic process, phasor measurement unit
\end{IEEEkeywords}

\vspace{-10pt}

\section{Introduction}
\vspace{-6pt}
%
The state estimator, a core component at the control room for monitoring and control, is subject to cyber attacks as measurement units are widely dispersed and communication networks transmitting measurements are vulnerable. It has been shown in \cite{Liu2009} that an adversary can stealthily distort the state estimation results 
without being detected by the bad data detection {(BDD), if the adversary has the full knowledge of system topological information including line admittance. 

Inspired by \cite{Liu2009}, many subsequent works \textcolor{black}{were }devoted to studying the false data injection attack (FDIA) models against state estimation \cite{DWd,Yuan2011a,Yang14,Rahman2012,Liu2015,Deng2019,Zhang2020,Esmalifalak2011,Yu2015}. Particularly, \textcolor{black}{in contrast to AC state estimation that is iterative and computationally harder to solve \cite{Liang2017}}, DC state estimation model is commonly used owing to \textcolor{black}{its} simplicity and linearity. \textcolor{black}{In \cite{DWd}, the authors proposed two different FDIAs targeted at pre-specified state variable or meter, respectively. In \cite{Yuan2011a}, a method to transform FDIA into a load distribution attack based on DC model is proposed. An FDIA method that selects the meters to manipulate aiming to cause the maximum damage can be found in \cite{Yang14}.}
\color{black}
The authors of \cite{Rahman2012} exploited  the concept of attacking region 
to design an FDIA model that requires only the line admittance and topology inside the attacking region. An efficient strategy to determine the optimal attacking region was later discussed in \cite{Liu2015}.  
In \cite{Deng2019}, the authors showed that an adversary can launch FDIA on a bus or superbus only if \textcolor{black}{the susceptance of every transmission line connected to that bus or superbus is known}. Despite important advancement towards relaxing the pre-knowledge of network information,  these methods still require partial information of a transmission network, which may still be unobtainable as the system model is critically protected by utilities. 

Recently, the authors of \cite{Zhang2020} proposed an FDIA model requiring no line parameters, yet the method enforces an assumption that the target bus has only one line connected to the outside, which may be rare, limiting its applications in practice. The independent component analysis (ICA) and the principal component analysis (PCA) have been applied in  \cite{Esmalifalak2011, Yu2015} to design FDIA against DC state estimation purely from measurements, without any grid topology and line parameters. However, these data-driven FDIA may not target specific states since the exact Jacobian matrix cannot be obtained. 



In this paper, we propose a novel FDIA model against DC state estimation that can target specific states without knowing any line parameters. Compared to network \color{black} topology \color{black} information critically protected and rarely transmitted, phasor measurement unit (PMU) measurements are easier to acquire through communication networks \textcolor{black}{or software supply chain attack between PMU vendors and customers (e.g., the SolarWinds attack \cite{article})}. Hence, we, from the viewpoint of an adversary, exploit only PMU data collected at the buses inside the attacking region to \textcolor{black}{estimate the line parameters of interest and} design a stealthy FDIA.
\textcolor{black}{In contrast to previous works that require measurements from remote terminal units (RTUs), the proposed attack model requires {only}} PMU measurements for \textcolor{black}{designing} the FDIA and can target specific states without any \textcolor{black}{pre-knowledge on} line parameters.

\color{black}
\vspace{-6pt}
\section{\color{black}DC State Estimation and Attacking Region \color{black}} \label{Background for the FDIA}
The power system state vector $\bm{x}$ to be estimated consists of voltage phasors of all buses, while the measurement vector $\bm{z}$ consists of power injections and line \textcolor{black}{flows}. 
The relationship between $\bm{z}$ and $\bm{x}$ can be formulated in a DC model:
\vspace{-5pt}
\small
\begin{equation}
\bm{z} = H\bm{x} + \bm{e}
 \vspace{-8pt}
\label{eq:relationship between z and x}
\end{equation}
\normalsize
where $\bm{e}$ denotes the vector of measurement error, which is typically assumed to be an independent Gaussian random vector; $H$ represents the measurement Jacobian matrix. 


To get the estimated state vector $\hat{\bm{x}}$, 
the  weighted least squares estimation problem is formulated  and solved as:
\vspace{-5pt}
\small
\begin{equation}
    \bm{\hat x} = {({H^T}WH)^{ - 1}}{H^T}W\bm{z}
    \label{eq:DC_state_estimation}
    \vspace{-5pt}
\end{equation}
\normalsize
where $W$ is a weight matrix.

For the detection of bad data, the residual-based BDD is typically used \cite{Liu2009}. The residual between the measurement and the estimated state vector is defined as $r = \left\| {\bm{z - }H\bm{\hat x}} \right\|_\infty$. Only if $r$ is smaller then a threshold value $\gamma $, the measurement $\bm{z}$ is regarded as normal without bad data or malicious data. 


Assuming an FDIA attack is launched by an adversary, the measurement vector after manipulation is 
${\bm{z}_{bad}} = \bm{z} + \bm{a}$, where $\bm{a}$ is termed as the attack vector. Let ${{{\bm{\hat x}}_{bad}}}=\hat{\bm{x}}+\bm{c}$ be the estimated states from the $\bm{z}_{bad}$. 
It has been shown in \cite{Rahman2012} that if $\bm{a}=H\bm{c}$, then the residual after the attack remains the same as that before the attack:
${r_{bad}}\bm{ = }\left\| {\bm{z}_{bad} -H\hat{\bm{x}}_{bad}}\right\|_{\infty}=\left\| {\bm{z} + \bm{a - }H(\bm{\hat x + c})} \right\|_{\infty}
= \left\| {\bm{z - }H\bm{\hat x} + \bm{a - }H\bm{c}} \right\|_{\infty} = r$, 
indicating that the attack can stealthily bypass the BDD without being detected. However, \textcolor{black}{it is} obvious that the adversary needs to know $H$, i.e., the entire network information (topology information and line parameters) of \textcolor{black}{the} power system, which is practically infeasible. 
Thus, many efforts (e.g., \cite{Rahman2012},\cite{Deng2019}) 
have been made to propose FDIA models using only partial network information or manipulating partial measurements. 

One effective method is to divide the system into the non attacking region and \textcolor{black}{the} attacking region. If the adversary knows the \textcolor{black}{topology} information \textit{inside} the attacking region and can manipulate line flow and power injection measurements, then 
\textcolor{black}{the} adversary can launch successful FDIA without being detected by the BDD\cite{Liu2009}. Network information or meter measurements \textit{outside} the attacking region are not required. 

In this paper, we also utilize the concept of attacking region to design the PMU-based FDIA. Nevertheless, the proposed FDIA requires no line parameters, irrespective of \textcolor{black}{whether the line is located} inside or outside the attacking region. 

\begin{figure}[!b]
\centering
\vspace{-20pt}
\includegraphics[width=0.4\textwidth,keepaspectratio=true,angle=0]{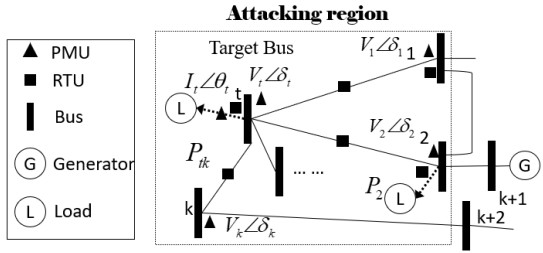}
\caption{The definition of attacking region}
\label{fig:attacking_region}
\vspace{-20pt}
\end{figure}

\vspace{-5pt}

\section{\color{black}{Constructing the FDIA Model Without Line Parameter Information} \color{black}}\label{construction for FDIA}


To construct the FDIA, we follow the common assumption that generator measurements cannot be attacked as they are well protected, while the measurements of line flow and power injection at load buses can be attacked because of the wide deployment of load meters and variation of load powers\cite{Liu2015}.


Assuming the adversary intends to attack a load bus $t$ that is termed as the target bus,  
then the attacking region defined in this paper consists of the \textcolor{black}{ following: the target bus, all buses} that have direct connection to the target bus as well as the transmission lines between these buses (see Fig. \ref{fig:attacking_region}). 


\textbf{Information needed in the attacking region:} The adversary needs to know: 1. which buses are connected to the target bus to determine the attacking region. 
Nevertheless,  the  \textcolor{black}{topology} information of other buses inside the attacking region (e.g., the line between bus 1 and bus 2 in Fig. \ref{fig:attacking_region}) is not needed. 2. similar to \cite{Liu2009,Deng2019}, 
voltage angles $\delta_i$, $\forall i \in {\Omega _A}$ where ${\Omega _A} = \{ 1,2,...,t,...,k\}$} denotes the set of bus numbers inside the attacking region. 
3. \color{black}{the current phasor measurement ${I_t\angle\theta_{t}}$ flowing from the target bus to the dynamic load, whereas any other line current measurements (e.g., ${I_{1t}\angle\theta_{1t},I_{2t}\angle\theta_{2t},...,I_{kt}}\angle\theta_{kt}$) are not needed by the adversary. Note that in contrast to previous works \cite{Liu2009, Yuan2011a, Rahman2012}, no RTU measurements containing power injections and line flows are needed by the adversary.} 



\color{black}
\textbf{Measurements to be manipulated 
in the attacking region:} To bypass the BDD of DC state estimation \cite{Rahman2012}, the adversary needs to manipulate all active bus power injections (e.g., $P_i$, $\forall i \in {\Omega _A}$) and all active line flows (e.g., $P_{tj},\forall j \in {\Omega _A},j \ne t$). Buses directly connected to  the  generator  buses cannot be attacked, because of the assumption that the adversary cannot manipulate generator measurements. 


\vspace{-5pt}

\subsection{Designing the Attack Vector}

\vspace{-2pt}
As discussed in Section \ref{Background for the FDIA}, in order to launch a stealthy FDIA to bypass the BDD, the attack vector $\bm{a}$ needs to satisfy \color{black} $\bm{a}=H\bm{c}$. \color{black}
More specifically, \textcolor{black}{if} the adversary wants the power operators to believe that the angle of \textcolor{black}{the} target node is $ {\tilde \delta _t}$ rather than the true value ${\delta _t}$, the adversary needs to manipulate the line flows and bus injections as follows:
\\
\noindent $\bullet$ substituting ${P_{tj}}$ by the fake ${\tilde P_{tj}}$ for $\forall j \in {\Omega _A},j \ne t$: 
\vspace{-5pt}
\small
\begin{equation}
{{\tilde P}_{tj}} = {B_{tj}}({{\tilde \delta }_t} - {\delta _j})
\vspace{-5pt}\label{fake_P_tj}
\end{equation}
\normalsize
\vspace{-5pt}
\noindent $\bullet$ substituting  ${P_t},{P_j}$ by the fake ${\tilde P_t},{\tilde P_j}$: \color{black}
\small
\begin{eqnarray}
{\tilde P_t} &=&\sum\nolimits_{j \in {\Omega _A}}^{j \ne t} {{{\tilde P}_{tj}}} 
\label{fake_P_t}\\
{{\tilde P}_j} &=& {P_j} + {{\tilde P}_{jt}} - {P_{jt}}
\label{fake_P_j}
\vspace{-15pt}
\end{eqnarray}
\normalsize
\color{black}
\noindent In other words, if the designed deviation $\bm{c}=[ \tilde {\delta}_t-\delta_t]^T$, then the attack vector $\bm{a}=[ \tilde{P}_{tj}-P_{tj},\tilde{P}_i-P_i]^T$, $\forall j\in\Omega_A,j \ne t, \forall i\in\Omega_A$. 
Obviously, the adversary needs to know accurate values of  $B_{tj}$ according to (\ref{fake_P_tj}). Nevertheless, line parameters are critically protected and cannot be easily acquired. In the next sections, we will 
draw upon intrinsic load dynamics and \textcolor{black}{the regression theorem of the Ornstein-Uhlenbeck (OU) process to estimate the line parameters $B_{tj}$ connected to the target bus $t$ and exploit the estimation result to} design the attack vector. 
\vspace{-5pt}
\subsection{\textcolor{black}{The Load Dynamics Inside the Attacking Region}}
\vspace{-2pt}
Inside an attacking region, the system can be represented by differential\textcolor{black}{-}algebraic equations:
\vspace{-5pt}
\small
\begin{eqnarray}
\bm{\dot x} &=& f(\bm{x,z})
\label{differentiable-algebraic eqn1}\\
\bm{z} &=& H\bm{x}
\label{differentiable-algebraic eqn2}
\vspace{-5pt}
\end{eqnarray}
\normalsize
where (\ref{differentiable-algebraic eqn1}) represents the load dynamics  and (\ref{differentiable-algebraic eqn2}) represents the relationship between the states and power injections as well as power flow relationship. 
It should be noted that only (\ref{differentiable-algebraic eqn2}) is considered in former FDIA works, whereas the dynamics described by (\ref{differentiable-algebraic eqn1}) are neglected by assuming that the system is in normal operating state. Nevertheless, we will show that the adversary can 
acquire essential information about physical systems\textcolor{black}{, such as line parameters,} by exploiting the load dynamics described by \eqref{differentiable-algebraic eqn1}. \color{black}
\vspace{-1pt}
For any bus $i$ inside the attacking region $\Omega_A$, the \textcolor{black}{stochastic} dynamic load model with DC power flow assumption can be represented as: 
\vspace{-5pt}
\small
\begin{eqnarray}
{\dot \delta _i} &=& \frac{1}{{{\tau _{{p_i}}}}}(P_i^s(1 + \sigma _i^p\xi _i^p) - {P_i})
\label{ddelta_dP}\\
{P_i} &=& \sum\limits_{j \in {\Omega _i}} {{B_{ij}}({\delta _i} - {\delta _j})}
\vspace{-20pt}
\label{bus reactive power injection}
\end{eqnarray}
\normalsize
\noindent where \textcolor{black}{ $\delta_{i}$ is the voltage angle of bus $i$; ${P_{i}}$ is the active power injection of bus $i$;  ${P_i^s}$ is the static active power injection of bus $i$; ${\Omega _i}$ is the set of buses connected to bus $i$; $B_{ij}$ is the susceptance between bus $i$ and $j$; $\tau _{{p_i}}$ is the active power time constant of bus $i$; ${\xi _i^p}$ is a standard Gaussian random variable; ${\sigma _i^p}$ describes the noise intensity of load variations.}  


This dynamic model can represent various loads 
including thermostatically controlled loads, induction motors, loads controlled by load tap changers, static loads, etc. in ambient conditions \cite{Caizares1995}. 
The difference among various \textcolor{black}{loads} is reflected by time constants, i.e., the relaxation rates of loads, which may range from 0.1s to 300s \cite{Karlsson1994, Wang2018}. 
Dynamic load models similar to (\ref{ddelta_dP})-(\ref{bus reactive power injection}) have also been proposed and used in previous works \cite{Pierrou2020a,Caizares1995,Mohammed2000} for stability analysis. 
In addition, 
\textcolor{black}{since} load powers are constantly varying in practice, 
we make the common assumption \cite{Caizares1995,Mohammed2000,Pierrou2020} that static load \textcolor{black}{powers} are perturbed by 
independent Gaussian noises, 
i.e., $P_i^s(1 + \sigma _i^p\xi _i^p)$.  

 Around steady state,\color{black}{ \eqref{ddelta_dP} can be linearized. By looking at \eqref{ddelta_dP} and \eqref{bus reactive power injection}, one can easily observe that the Jacobian matrix $J_{\bm{P\delta}}=\frac{\partial\bm{P}}{\partial\bm{\delta}}$ required for linearization carries crucial information of line parameters, as  $\frac{{\partial {P_i}}}{{\partial {\delta _i}}} = \sum\limits_{j \in {\Omega _i}} {{B_{ij}}} $  and  $\frac{{\partial {P_i}}}{{\partial {\delta _j}}}={\rm{ - }}{B_{ij}}$, for $i\neq j$. Therefore, the linearized load dynamics inside the attacking region can be described in a matrix form as follows:
\vspace{-7pt}
\begin{equation}
\label{swing-load-matrix1}
\scriptsize
\begin{array}{l}
\left[ {\begin{array}{*{20}{c}}
{{{\dot \delta }_1}}\\
{...}\\
{{{\dot \delta }_k}}
\end{array}} \right] = \underbrace {\left[ {\begin{array}{*{20}{c}}
{\frac{{ - 1}}{{{\tau _{{p_1}}}}}}&{...}&0\\
0&{...}&0\\
0&0&{\frac{{ - 1}}{{{\tau _{{p_k}}}}}}
\end{array}} \right]\left[ {\begin{array}{*{20}{c}}
{\sum\limits_{j \in {\Omega _i}} {{B_{1j}}} }&{...}&{ - {B_{1k}}}\\
{...}&{...}&{...}\\
{ - {B_{k1}}}&{...}&{\sum\limits_{j \in {\Omega _i}} {{B_{kj}}} }
\end{array}} \right]}_A\left[ {\begin{array}{*{20}{c}}
{{\delta _1}}\\
{...}\\
{{\delta _k}}
\end{array}} \right]\\
 + \underbrace {\left[ {\begin{array}{*{20}{c}}
{\frac{1}{{{\tau _{{p_1}}}}}}&{...}&0\\
0&{...}&0\\
0&0&{\frac{1}{{{\tau _{{p_k}}}}}}
\end{array}} \right]\left[ {\begin{array}{*{20}{c}}
{P_1^s}&{...}&0\\
0&{...}&0\\
0&{...}&{P_k^s}
\end{array}} \right]\left[ {\begin{array}{*{20}{c}}
{\sigma _1^p}&{...}&0\\
0&{...}&0\\
0&{...}&{\sigma _k^p}
\end{array}} \right]}_S\left[ {\begin{array}{*{20}{c}}
{\xi _1^p}\\
{...}\\
{\xi _k^p}
\end{array}} \right]
\end{array}
\small
\end{equation}
\normalsize
Equation \eqref{swing-load-matrix1} can be written in the following compact form:
\vspace{-10pt}

\color{black}
%
\small
\begin{equation}
\label{eq:OU}
\dot{\bm{\delta}}=A\bm{\delta}+S\bm{\bm{\xi }^p}
\end{equation}
\normalsize
from which it can be observed that $\bm{\delta}$ is a multi-dimensional \textcolor{black}{OU} process \cite{Gardiner2009} that is Markovian and Gaussian. 
\textcolor{black}{In addition, as seen from \eqref{swing-load-matrix1},  
the} system state matrix $A$ contains important information of \textcolor{black}{the} time constants and the Jacobian matrix, from which the line parameters can be extracted. In the next sections, a method to estimate matrix $A$ purely from PMU measurements will be introduced. \textcolor{black}{Particularly, the method enables the adversary to estimate the line parameters $B_{tj}$ connected to the target bus $t$} and build the attack vector for launching FDIA \textcolor{black}{without requiring any other information other than the available PMU data within the attacking region}. 



\vspace{-5pt}
\subsection{\color{black}{The Regression Theorem for the OU Process}}
\label{3d}

Considering the multi-dimensional OU process in (\ref{eq:OU}), if the system state matrix $A$ is stable, which is satisfied in normal operating state, it can be shown that the $\tau$-lag correlation matrix of $\bm{\delta}$ satisfies a differential equation \cite{Gardiner2009}:
\vspace{-5pt}
\small
\begin{equation}
\frac{d}{d\tau}\left[C(\tau)\right]=AC(\tau)
\label{eq:regressiontheorem}
\vspace{-5pt}
\end{equation}
\normalsize
which is termed as the regression theorem of \textcolor{black}{the} multi-dimensional OU process. Particularly, the $\tau$-lag correlation matrix of $\bm{\delta}$ is defined as:
\small
\begin{equation}
\begin{array}{l}
C(\tau ) = E[({\bm{\delta}_{t + \tau }} - E[{\bm{\delta}_{t }}]){({\bm{\delta}_t} - E[{\bm{\delta}_t}])^T}]\\
\end{array}
\label{eq:covariance}
\end{equation}
\normalsize
where $E[.]$ denotes the expectation operator. 



We can therefore estimate $A$ from the $\tau$-lag correlation matrix of $\bm{\delta}$ by solving (\ref{eq:regressiontheorem}): 
\small
\vspace{-5pt}
\begin{equation}
A =  \frac{1}{\tau }\ln [C(\tau )C{(0)^{ - 1}}]
\label{eq:estimation_of_matrix_A_tau}
\vspace{-5pt}
\end{equation}
\normalsize
Equation (\ref{eq:estimation_of_matrix_A_tau}) provides an ingenious way of estimating the system state matrix $A$ that carries significant physical system information purely \textcolor{black}{from the statistical properties} of PMU measurements. In practical \textcolor{black}{applications}, the $\tau$-lag correlation matrix $C(\tau)$ can be estimated by the sample correlation matrix obtained from \textcolor{black}{a finite amount of PMU} data as below:  
\vspace{-7pt}
\small
\begin{equation}
{{\bm{\hat \mu }}_\delta } = \frac{1}{N}\sum\limits_{i = 1}^N {{\bm{\delta }^{(i)}}}    \\
\label{mu_x}
\vspace{-6pt}
\end{equation}
\begin{equation}
\hat C(0) = \frac{1}{{N - 1}}\sum\limits_{i = 1}^N {[({\bm{\delta }^{(i)}} - {{\bm{\hat \mu }}_\delta }){{({\bm{\delta }^{(i)}} - {{\bm{\hat \mu }}_\delta })}^T}]} 
\end{equation}
\label{C_zero}
\vspace{-6pt}
\begin{equation}
\hat C(\Delta t) = \frac{1}{{N - M - 1}}\sum\limits_{i = 1 + M}^N {[({\bm{\delta }^{(i)}} - {{\bm{\hat \mu }}_\delta }){{({\bm{\delta }^{(i - M)}} - {{\bm{\hat \mu }}_\delta })}^T}]}  
\end{equation}
\label{C_t}
\normalsize
\noindent where  $N$ is the sample size of voltage angles, $M$ is the number of samples that corresponds to the selected time lag $\Delta t$, ${\bm{\delta^{(i)}}} = {[\delta_1^{(i)},...,\delta_t^{(i)},...,\delta_k^{(i)}]^T}$ represents the ${i^{th}}$ measurement of all voltage angles in the attacking region. 
Thus, the adversary can obtain the estimated matrix $\hat A$ as follows:
\small
\begin{equation}
{\hat{A}} =  \frac{1}{\Delta t }\ln [\hat{C}(\Delta t )\hat{C}{(0)^{ - 1}}]
\label{eq:estimation_of_matrix_A_eqn}
\end{equation}
\normalsize

\subsection{Estimation \textcolor{black}{of} Time Constants and Line Parameters} 

\label{3e}
Once $\hat{A}$ is calculated from the collected PMU data through (\ref{eq:estimation_of_matrix_A_eqn}), the adversary can apply the least-squares estimation (LSE) optimization formulation 
to estimate the time \textcolor{black}{constant $\tau_{p_t}$ and finally extract the desired line parameters for the attack vector}. To this end,  \eqref{ddelta_dP} can be re-written in its discrete form:
\small
\begin{equation}
\label{least_square_estimation_for_time_constant}
\underbrace {\left[ \begin{array}{l}
\frac{1}{{\Delta t}}(\delta _t^{(2)} - \delta _t^{(1)})\\
...\\
\frac{1}{{\Delta t}}(\delta _t^{(n)} - \delta _t^{(n - 1)})
\end{array} \right]}_{\bm{Y}} = \underbrace {\left[ \begin{array}{l}
{\hat \mu _{{P_t}}} - P_t^{(1)}\\
...\\
{\hat \mu _{{P_t}}} - P_t^{(n - 1)}
\end{array} \right]}_{\bm{X}} \left[ {\frac{1}{{{\tau _{{p_t}}}}}} \right]
\vspace{-10pt}
\end{equation}
\normalsize
where $n$ is the sample size,  $P_t^{(i)}$ is the ${i^{th}}$ observation of active power of the target bus $t$, $i=1,...,n$; ${\hat{\mu} _{{P_t}}}$ denotes the sample mean of $P_t^{(i)}$; $\delta _t^{(i)}$ represents the ${i^{th}}$ observation of voltage angle of bus $t$. 
Therefore, the time constant can be estimated \color{black}as $1/\hat{\tau}_{p_t} = {({\bm{X}^T}\bm{X})^{ - 1}}{\bm{X}^T}\bm{Y}$ \color{black} by the LSE method.  Note that the real power injection at bus $t$ can be computed from the PMU measurements of voltage and current phasors at bus $t$. 

Once ${\hat{\tau}_{{p_t}}}$ is obtained, \textcolor{black}{the adversary can use the result together with the estimated $\hat{A}$ from (\ref{eq:estimation_of_matrix_A_eqn}) to estimate the line parameters $\hat B_{tj}$}. The proposed FDIA model against DC state estimation without pre-knowledge of line parameters is summarized in \textbf{Algorithm 1} below. 


\vspace{-3pt}

\textbf{Remarks:}\\

\vspace{-10pt}

\noindent $\bullet$ \textcolor{black}{In this paper, $300$s emulated PMU measurements with a sampling rate of $60$ Hz are used for Step 2, while only a few milliseconds e.g., 0.0041s, are needed for Steps 3-6.}



\noindent $\bullet$ The proposed FDIA against DC state estimation needs only a small number of PMU measurements inside the attacking region, whereas no additional measurements (i.e., line flows and line parameters) are required to carry out the attack.  
\\
\noindent \color{black} $\bullet$ The adversary is aware of the network connectivity, which is a reasonable assumption in many works \cite{Rahman2012,Liu2014,Deng2019,Zhang2020}. Yet, the exact line parameters that may vary in practice are not required, highlighting the superiority of the proposed method compared to topology dependent methods.  \color{black}

\vspace{-5pt}
\begin{algorithm}[!ht]
\caption{The proposed FDIA model against DC state estimation without line parameters} 
\label{Flow chart of the FDIA} 
\begin{algorithmic} 
\State 
\hspace{-10pt}\textbf{1}. Choose one target bus $t$ and determine attacking region $\Omega_A$. \\ 
\hspace{-10pt}\textbf{2}. Collect $N$ 
voltage measurements for $k$ buses inside attacking region and $n$ 
current phasor measurements only for bus $t$ from PMUs; Calculate the 
real power injections at bus $t$.\\
\hspace{-10pt}\textbf{3}. Calculate {the estimated matrix $\hat{A}$ through \eqref{mu_x}-\eqref{eq:estimation_of_matrix_A_eqn}}. \\
\hspace{-10pt}\textbf{4}. Estimate the time constants ${{\hat \tau }_{{p_t}}}$ through \eqref{least_square_estimation_for_time_constant}.\\  
\hspace{-10pt}\textbf{5}. Estimate the line admittance $\hat{B}_{tj}$ 
from $\hat A$ and  ${{\hat \tau }_{{p_t}}}$ . \\
\hspace{-10pt}\textbf{6}. Given the target malicious phasor ${\tilde \delta _t}$, design the attack vector $\bm{a}$  through \textcolor{black}{\eqref{fake_P_tj}-\eqref{fake_P_j}}. 
\end{algorithmic}
\end{algorithm}
\vspace{-18pt}
\subsection{Analysis on the Performance of the Proposed FDIA Model}
\vspace{-3pt}
In this section, sufficient conditions for \textbf{Algorithm 1} \textcolor{black}{are} presented as the theoretical basis for the proposed FDIA.  

\noindent\textit{Theorem 1}: If the number of independent measurements is equal to the number of states to be estimated, then the proposed algorithm is perfect, i.e., the residual remains the same before and after the attack.\\ 
 \textit{Proof:} 
the Jacobian matrix estimated by the adversary can be represented by \small$\hat H = H + \Delta H$ \normalsize due to the estimation error in $B_{tj}$.  Then the attack vector designed by the adversary takes the form \small$\bm{a} = \hat H c=(H + \Delta H)\bm{c}$\normalsize, where $\bm{c}$ is the deviation intended to be injected to bus $t$. The measurement to be manipulated is 
\small
$\bm{z}_{bad}=\bm{z}+\bm{a}=\bm{z} + (H + \Delta H)\bm{c}\label{eq:zbad}$. 
\normalsize
At the power operators' side, if DC state estimation is exploited, the result of \eqref{eq:DC_state_estimation} is 
\small
${{\bm{\hat x}}_{bad}} = \bm{\hat x + c} + {({H^T}WH)^{ - 1}}H^TW\Delta H\bm{c}$.
\normalsize
Consequently, 
the residual after the FDIA is \small
${r_{bad}} = \left\| {{\bm{z}_{bad}} - H{{\bm{\hat x}}_{bad}}} \right\|_{\infty}
 = \left\| {r + (I - H{{({H^T}WH)}^{ - 1}}{H^TW})\Delta H\bm{c}} \right\|_{\infty}$.
\normalsize
\textcolor{black}{It can be observed that if the number of independent measurements is the same as that of states to be estimated, i.e., $H$ is a square matrix with full rank, then the FDIA is perfect as $r_{bad}=r$.}

\textcolor{black}{However in real-life applications, the number of measurements is typically larger than that of states. If the matrix $H$ is full rank but not a square matrix, the residual after the attack may not remain the same and depends on $\Delta H \bm{c}$. }

It will be demonstrated in Section \ref{casestudy} that the proposed FDIA can launch targeted large deviation attacks with a high success rate owing to the good estimation accuracy of $\hat{H}$. \color{black}



\color{black}
\vspace{-5pt}
\section{Case Studies}\label{casestudy}
\vspace{-3pt}
In this section, the proposed FDIA model is tested on a modified IEEE 39-bus 10-generator system. 
\color{black}
We consider the worst situation for the adversary, i.e, the power system is fully measured by RTUs for state estimation purposes at the control room, 
while the adversary can only collect limited PMU data 
within the attacking region. 
\textcolor{black}{In this study,} there are totally 85 RTU \color{black} measurements containing 39 bus injections and 46 line flows used by the system operator for state estimation. All 85 measurements contain independent Gaussian noises described by 
$\forall e_i \in \bm{e}$ , ${e_i}\sim {\rm{ }}\mathcal{N}(0,{0.05^2})$ in \eqref{eq:relationship between z and x}. \textcolor{black}{Note that these RTU measurements are not used by the adversary.} 

\color{black}
\vspace{-5pt}
\subsection{Constructing the Proposed FDIA without Line Parameters}\label{sectionexampleI}
\vspace{-5pt}
Assuming the \textcolor{black}{adversary} intends to manipulate the voltage angle of bus 15, then the attacking region includes bus 15 as well as buses 14 and 16 that are directly connected to bus 15, as \textcolor{black}{defined} in Section \ref{construction for FDIA} \textcolor{black}{and} shown in Fig. \ref{False data attack in IEEE39 system}.  
The time constants of the loads inside the attacking region are set arbitrarily as $[{\tau _{{p_{14}}}},{\tau _{{p_{15}}}},{\tau _{{p_{16}}}}] = [{\rm{0}}{\rm{.1}},{\rm{27}}{\rm{.
88}},{\rm{206}}{\rm{.01}}]$; the process noise intensities $\sigma_i^p$, $i\in\{14,15,16\}$  in \eqref{ddelta_dP} describing power variations are set to be 1. 
As the proposed FDIA is purely measurement-based, its robustness against measurement noise should be tested. Following the approach in \cite{Zhou2013}, PMU measurement noises with a standard deviation of 10\% of the largest state changes between discrete time \textcolor{black}{steps} are added to the PMU data of voltages and currents utilized by the adversary.  \textcolor{black}{
It will be shown that through the proposed FDIA method, the adversary can stealthily bypass BDD to manipulate the angle of bus 15.} 
\begin{figure}[!b]
\centering
\vspace{-20pt}
\includegraphics[width=0.28\textwidth,keepaspectratio=true,angle=0]{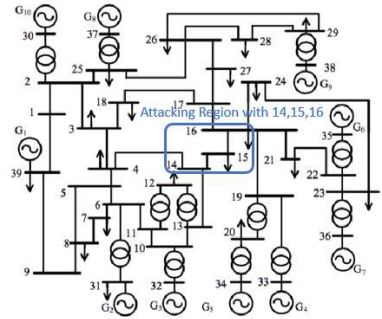}
\vspace{-5pt}
\caption{The attacking region for bus 15.}
\label{False data attack in IEEE39 system}
\end{figure}

By \textbf{Algorithm 1}, the adversary collects $300$s, 60 Hz PMU data of 
voltage angle measurements for all the buses inside the attacking region (i.e., bus 14, 15, and 16) as well as current phasor measurements of the target bus 15, 
i.e., $N=18000$ in \eqref{mu_x} and $n=10$ in \eqref{least_square_estimation_for_time_constant}, 
then the matrix $A$ and the time constants $1/{\tau _{{p_{15}}}}$ can be estimated by \eqref{eq:estimation_of_matrix_A_eqn}-\eqref{least_square_estimation_for_time_constant}. 
With these estimations, \textcolor{black}{the estimated line admittances  $\hat{B}_{14 - 15}, \hat{B}_{16 - 15}$} can be extracted. Table \ref{Regression Theorem for OU process} presents the estimation results for the time constants and line admittances, showing relatively good estimation accuracy. 

\begin{table}[!b]
\centering
\vspace{-11pt}
\captionsetup{justification=centering}
  \caption{\textsc{True and Estimated Values of Line Parameters and Time Constants }}
  \label{Regression Theorem for OU process}
\begin{tabular}{|c|c|c|}
\hline
\textbf{Time constant}& \textbf{True value (s)}& \textbf{Estimated value (s)}\\
 \hline
$\tau_{p_{15}}$ & 27.88 & 27.87 
\\
\hline
\textbf{Line Parameter}& \textbf{True value (pu)}& \textbf{Estimated value (pu)}\\
\hline
\begin{tabular}{c} ${B_{14 - 15}}$ \end{tabular} & -45.786 & -39.634
\\
\hline
\begin{tabular}{c} ${B_{16 - 15}}$ \end{tabular}& -105.42 &	-101.64
\\
\hline
  \end{tabular}
  \vspace{-5pt}
\end{table}
\normalsize

Once the line admittances are estimated, the adversary can 
design the attack vector 
according to  \textcolor{black}{\eqref{fake_P_tj}-\eqref{fake_P_j}}.  
Assuming the adversary launches an attack such that $\tilde{\delta}_{15}=10^\circ$, Table \ref{voltage magnitude and angle before and after attack_region_DC} presents a comparison between the estimated state values inside the attacking region through the DC state estimation before and after the attack, which validates the effectiveness of the proposed FDIA method.
  
\begin{table}[!tb]
\centering
\vspace{-5pt}
\captionsetup{justification=centering}
  \caption{\textsc{DC State Estimation Results Before and After the Attack ${\tilde \delta_{15}} =  10^ \circ$}}\label{voltage magnitude and angle before and after attack_region_DC}
\begin{tabular}{|c|c|c|}
\hline
\textbf{State Variables}& \textbf{Before attack (degree)}& \textbf{After attack (degree)}\\
\hline
\begin{tabular}{c} ${ \delta_{14}} $ \end{tabular} & $2.8300^ \circ$ & $2.8720^ \circ$
\\
\hline
\begin{tabular}{c} ${ \delta_{15}} $ \end{tabular} & $2.4557^ \circ$ & $9.5460^ \circ$
\\
\hline
\begin{tabular}{c} ${ \delta_{16}} $ \end{tabular} & $4.0061^ \circ$ & $3.9344^ \circ$
\\
\hline
  \end{tabular}
  \vspace{-5pt}
\end{table}
\normalsize
\vspace{-5pt}
\subsection{Different Attack Vectors under DC State Estimation}\label{sectionexampleII}
\vspace{-5pt}
As shown in \eqref{fake_P_tj}, the estimation error of \textcolor{black}{the} line admittance $B_{tj}$ may affect the residual after the attack. 
To test the performance of the FDIA model under the estimation error of line parameters, different attacks are tested. Various Monte Carlo simulations using 1000 samples have been carried out. 
\begin{figure} [!b]
\vspace{-20pt}
\centering
\subfigure[]
{ 
\begin{minipage}[t]{0.5\linewidth}
\centering
\includegraphics[width=1.65in]{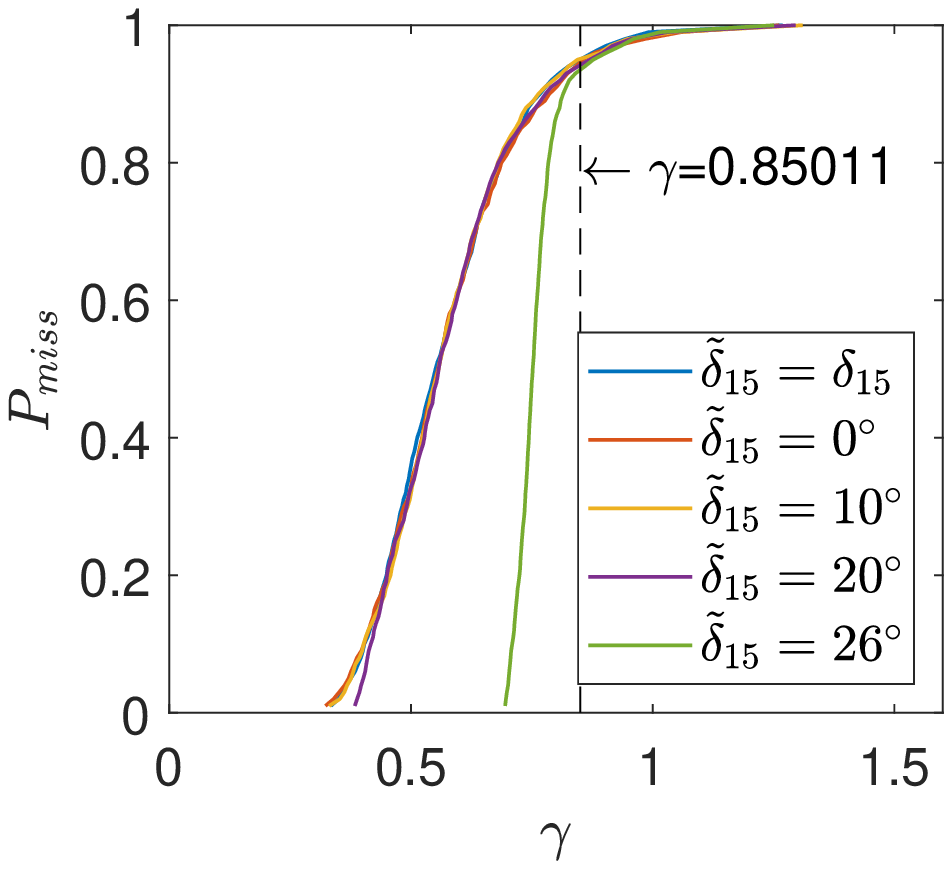}
\vspace{-100pt}
\end{minipage}%
}%
\subfigure[]
{ 
\begin{minipage}[t]{0.5\linewidth}
\centering
\includegraphics[width=1.65in]{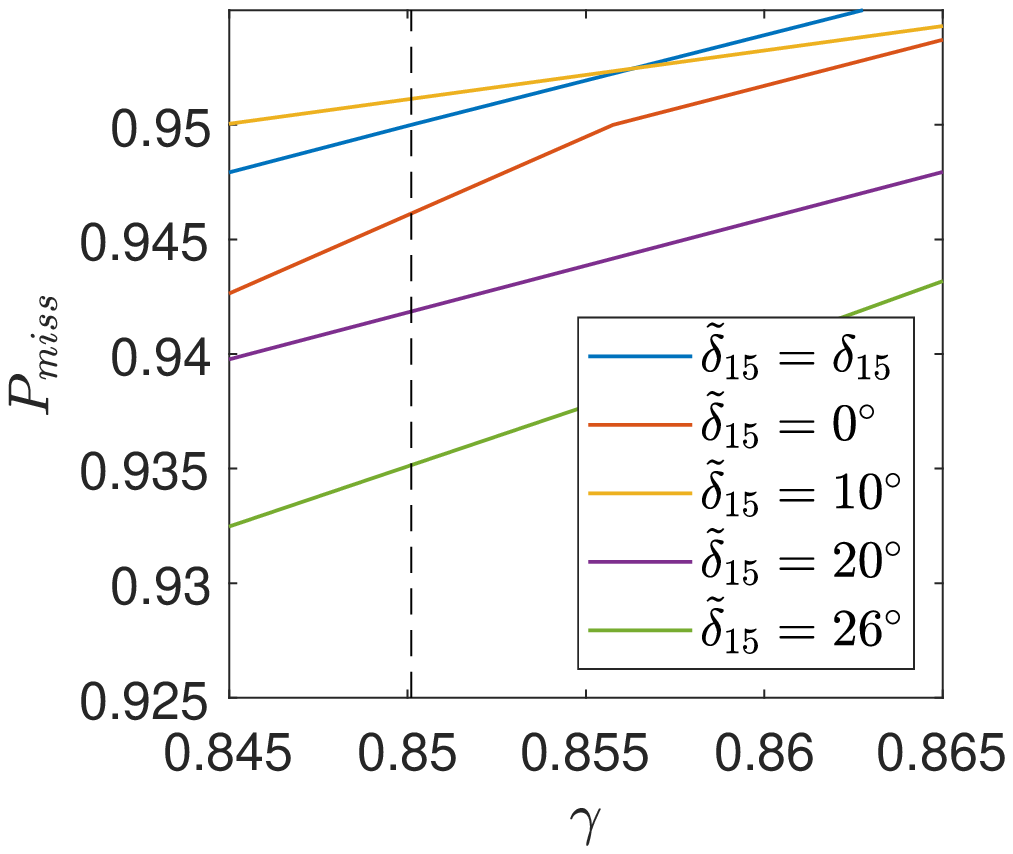}
\vspace{-100pt}
\end{minipage}%
}%
\centering
\vspace{-10pt}
\caption{(a). Probability to pass the BDD for different attacks if DC state estimation is applied; (b). Zoomed-in picture of (a). }
\label{different_angle_attacks_DC} 
\end{figure}

Fig. \ref{different_angle_attacks_DC} presents the probabilities of the FDIAs to bypass the BDD for different attacks. 
If we choose 95\%-quantile of the residual before the attacks, i.e, $\gamma= 0.85011$ to be the BDD threshold, the probabilities to bypass the BDD for ${\tilde \delta_{15} = 0 ^\circ }$,${\tilde \delta_{15} = 10 ^\circ }$, ${\tilde \delta_{15} = 20 ^\circ }$, ${\tilde \delta_{15} = 26 ^\circ }$ are 94.7\%, 95.3\%, 94.2\% and 93.5\%, respectively, showing that the adversary can successfully bypass the BDD even for big deviation attacks. 
\vspace{-8pt}
\subsection{Different Attack Vectors under AC State Estimation}\label{sectionexampleIII}
\vspace{-5pt}
Although the proposed FDIA 
is based on DC model, we also test its performance 
if AC state estimation is implemented in the control room. As seen in Table \ref{voltage magnitude and angle before and after attack_region_AC}, 
the attack $\tilde {\delta}_{15}=10^\circ$ can still be launched successfully. Meanwhile, the states of other buses inside the attacking region are almost unaffected.  
  
\begin{table}[!b]
\centering
\vspace{-5pt}
\captionsetup{justification=centering}
  \caption{\textsc{AC state estimation results before and after attack with ${\tilde \delta_{15}} =  10^ \circ$}}\label{voltage magnitude and angle before and after attack_region_AC}
\begin{tabular}{|c|c|c|}
\hline
\textbf{State Variables}& \textbf{Before attack}& \textbf{After attack}\\
\hline
\begin{tabular}{c} ${ V_{14}\angle\delta_{14}} $ \end{tabular} & $0.9330\angle 3.2361^ \circ$ & $0.9346\angle 3.1980^ \circ$
\\
\hline
\begin{tabular}{c} ${ V_{15}\angle\delta_{15}} $ \end{tabular} & $0.9385\angle 2.6828^ \circ$ & $0.9487\angle 10.4593^ \circ$
\\
\hline
\begin{tabular}{c} ${ V_{16}\angle\delta_{16}} $ \end{tabular} & $0.9594\angle 4.1889^ \circ$ & $0.9589\angle 4.2063^ \circ$
\\
\hline
  \end{tabular}
  \vspace{-23pt}
\end{table}


Nevertheless, if the designed deviation $\bm{c}$ becomes larger, the proposed FDIA method may fail as the adversary does not manipulate reactive power in the system due to the assumption of DC model. As presented in Fig. \ref{different_angle_attacks}, 
the probabilities to bypass the BDD in AC state estimation for ${\tilde \delta_{15} = 0 ^\circ }$,${\tilde \delta_{15} = 10 ^\circ }$, ${\tilde \delta_{15} = 15 ^\circ }$, ${\tilde \delta_{15} = 20 ^\circ }$ are 95.5\%, 93.6\%, 0\% and 0\%, respectively, if \textcolor{black}{the} 95-quantile of the residual in AC state estimation before the attacks, i.e., $\gamma= 0.86936$, is used. These results \textcolor{black}{show} that, \textcolor{black}{if the adversary intends to launch large deviation attacks with higher success rates,} future efforts on designing targeted FDIA attacks against AC state estimation \textcolor{black}{are} needed. 
\vspace{-10pt}
\begin{figure}[!t]
\vspace{-15pt}
\centering
\subfigure[]
{ 
\begin{minipage}[t]{0.5\linewidth}
\vspace{-15pt}
\centering
\includegraphics[width=1.65in]{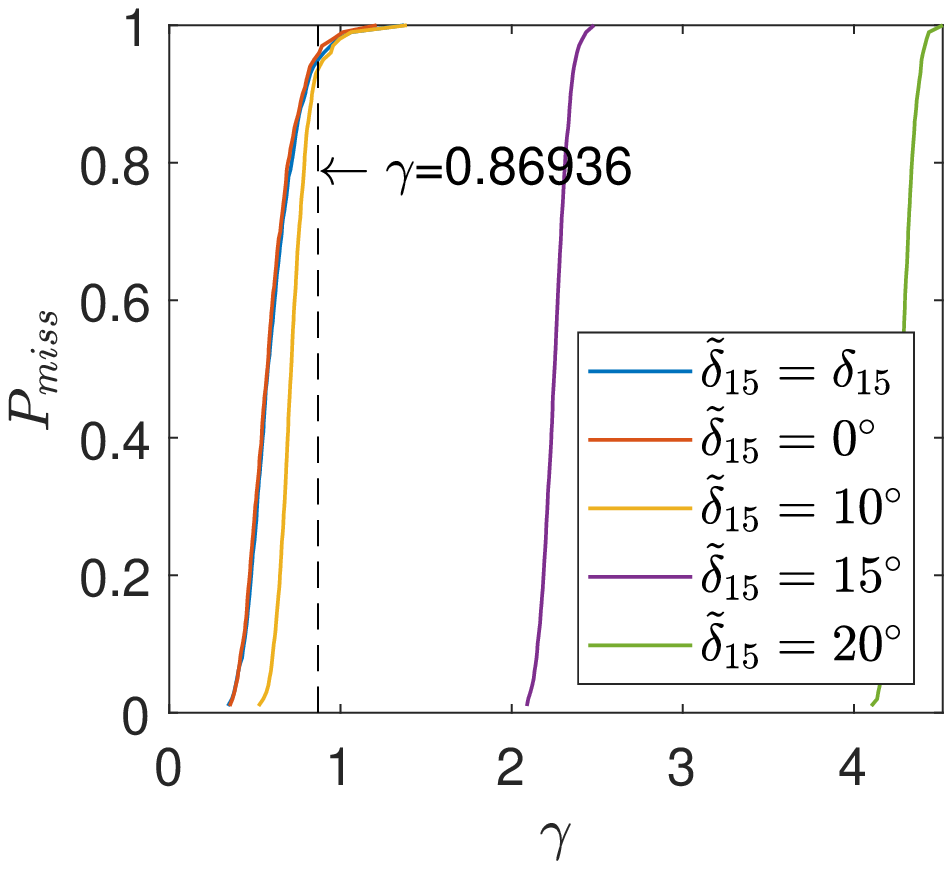}
\end{minipage}%
}%
\subfigure[]
{
\begin{minipage}[t]{0.5\linewidth}
\vspace{-15pt}
\centering
\includegraphics[width=1.65in]{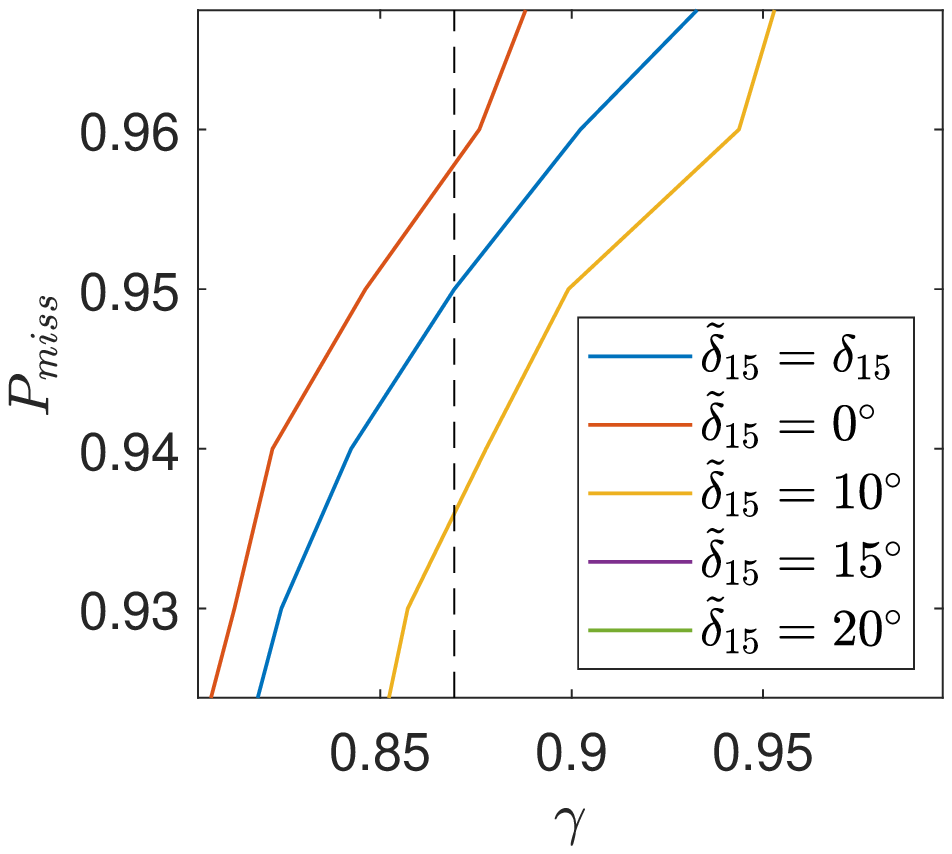}
\end{minipage}%
}
\centering

\caption{(a). Probability to pass the BDD for different attacks if AC state estimation is applied; (b). Zoomed-in picture of (a).\vspace{-15pt}}
\label{different_angle_attacks} 
\vspace{-5pt}
\end{figure}
\vspace{-12pt}
\section{Conclusion}
\vspace{-3pt}
In this paper, a novel FDIA model against DC state estimation \textcolor{black}{using PMU data} has been proposed, which can target specific states without any \textcolor{black}{pre-knowledge of} line parameters. Sufficient conditions for the proposed FDIA model are also provided. Numerical results showed that the proposed model can launch targeted large deviation attacks with high probabilities, if DC model is exploited. Future works on designing FDIA against AC state estimation are needed \textcolor{black}{for large attacks}. 



\medskip
\bibliographystyle{IEEEtran}
\vspace{-15pt}
\bibliography{IEEEabrv,My_Collection_with_no_color.bib}

\end{document}